# A Wave of Resignations in the Aftermath of Remote Onboarding

Darja Šmite[a,*], Franz Zieris[a] and Lars-Ola Damm[b]

[a]*Blekinge Institute of Technology, Karlskrona, Sweden*
[b]*Ericsson, Karlskrona, Sweden*

ABSTRACT

**Context:** The COVID-19 pandemic has permanently altered workplace structures, normalizing remote work. However, critical evidence highlights challenges with fully remote arrangements, particularly for software teams. **Objective:** This study investigates employee resignation patterns at Ericsson, a global developer of software-intensive systems, before, during, and after the pandemic. **Method:** Using HR data from 2016-2025 in Ericsson Sweden, we analyze how different work modalities (onsite, remote, and hybrid) influence employee retention. **Results:** Our findings show a marked increase in resignations from summer 2021 to summer 2023, especially among employees with less than five years of tenure. Employees onboarded remotely during the pandemic were significantly more likely to resign within their first three years, even after returning to the office. Exit surveys suggest that remote onboarding may fail to establish the necessary organizational attachment, the feeling of belonging and long-term retention. By contrast, the company's eventual successful return to pre-pandemic retention rates illustrates the value of differentiated work policies and supports reconsidering selective return-to-office (RTO) mandates. **Conclusions:** Our study demonstrates the importance of employee integration practices in hybrid environments where the requirement for in-office presence for recent hires shall be accompanied by in-office presence from their team members and more senior staff whose mentoring and social interactions contribute to integration into the corporate work environment. We hope these actionable insights will inform HR leaders and policymakers in shaping post-pandemic work practices, demonstrating that carefully crafted hybrid models anchored in organizational attachment and mentorship can sustain retention in knowledge-intensive companies.

## 1. Introduction

The widespread adoption and popularity of remote work during and after the COVID-19 pandemic has profoundly influenced the rules for employee attraction and retention [9, 22]. A substantial body of research suggests that remote work opportunities are positively associated with job satisfaction and reduced attrition [2, 3, 11, 12, 17]. With remote work increasingly normalized, especially in the tech sector, even limited flexibility, such as two to three days of office presence per week, can significantly influence employees' willingness to remain in their current job. Workers have reported readiness to accept sizable pay cuts in return for the option to work remotely [2], making remote work a key attractor in recruitment efforts. Some researchers even argue that employees strive to spend as little time in the workplace as possible [17].

However, despite its benefits, remote work is also infamous for its challenges and paradoxical effects (positive and negative at the same time) on software engineers' experiences, including ability to focus, motivation, quality of meetings, social connections, and work-life balance [10]. These dual effects of remote work are particularly relevant in the context of the "Great Resignation", a record wave of voluntary quits observed between 2021 and 2022. According to the U.S. Job Openings and Labor Turnover Survey,[1] resignations rates during this period reached historic highs, with similar patterns observed internationally (in Europe,[2] France,[3] or the UK[4]). Some suggest that the resignation is connected with the pushback expressed by those who have had better-than-expected experiences with remote work during the pandemic. The results of the Survey of Work Arrangements and Attitudes [1] covering American workers confirm that full-time return-to-office (RTO) mandates motivate 6% of respondents to quit their jobs and further 36% to seek a new job. Therefore, many workers are said to be re-sorting across employers and into jobs that better suit their preferences for work modality. There are other potential explanations for the employment changes. A qualitative analysis of the changes in work- and quit-related online posts on Reddit [20] suggests that the reasons for resignations are likely linked with mental health and work-related distress that prevailed since the onset of the pandemic. Finally, some further speculate that the eventual decline in resignations rates post-2022 may be partially attributed to growing economic uncertainty and recessionary pressures.

This study was funded by the Swedish Knowledge Foundation through the projects S.E.R.T. (grant number 2018/010) and WorkFlex (grant number 2022/0047).

*Corresponding author

✉ darja.smite@bth.se (D. Šmite); franz.zieris@bth.se (F. Zieris); lars-ola.damm@ericsson.com (L. Damm)
ORCID(s): 0000-0003-1744-3118 (D. Šmite); 0000-0001-7482-6022 (F. Zieris)

---

[1]U.S. Bureau of Labor Statistics, Quits: Total Nonfarm [JTSQUR], Federal Reserve Bank of St. Louis: https://fred.stlouisfed.org/series/JTSQUR
[2]Job vacancy statistics by NACE Rev. 2 activity, Eurostat: https://data.europa.eu/data/datasets/hj5vu9sfjbp2qkoewhzoa
[3]Labor movements (les mouvements de main-d'œuvre), Dares: https://dares.travail-emploi.gouv.fr/donnees/les-mouvements-de-main-doeuvre
[4]UK job to job resignations, Statista: https://www.statista.com/statistics/1283657

One may notice the paradox that the highest wave of resignations coincides with the peak of fully remote working. This raises an important question: How come remote work that increases job satisfaction also leads to people quitting? Of equal importance is the understanding of who is likely to quit and why.

## 2. Background and Related Work

The paradox of remote work that appears to offer both higher job satisfaction and an increased risk of voluntary attrition is evident in related research on this topic. On the surface, research associated remote work with positive effects on both job satisfaction and retention [2, 3, 11, 12, 16, 17]. However, these pre-pandemic studies often focus on limited remote work opportunities. Critical studies report that remote work can also increase turnover intentions [14]. Further, comprehensive studies reveal that job satisfaction increases with remote work up to a point, but excessive remote work may reduce it, forming an inverted U-shaped relationship [12].

The known reasons for resignations in relation to remote working during and after the pandemic vary.

**Return-to-office (RTO) mandates** were found to be blamed for the abnormally high employee turnover in an extensive analysis of the records of over three million tech and finance workers' employment histories from LinkedIn [7]. Vulnerable groups included females, seniors, and more skilled employees. Additionally, companies with RTO mandates faced more challenges filling their job vacancies, implying that the corporate attractiveness as an employer is significantly damaged by the changes in work regulation.

**Work-related distress** surfaced as one of the key reasons for quitting in an analysis of Reddit posts [20]. Many posts reflected the feelings of being lost, stressed, anxious and overwhelmed at work, highlighting the emotional burden of prolonged remote work [20].

**Erosion of attachment and belonging** was linked with remote work in several post-pandemic studies. Eng et al. [9] interviewed Swedish SME managers who described increasing challenges in maintaining employee loyalty. Many cited weakened interpersonal connections and toward the company due to remote work, which ultimately made it harder to retain staff [9]. De Souza Santos and Ralph [23] report changes in teamwork quality, coordination problems, and gradual detachment among teammates in remote-first and hybrid software teams. Similarly, Tkalich et al. [25] report that a lack of spontaneous interaction in remote and hybrid teams significantly hinders teams, leaves remoters to work in isolation without frequent feedback, and reduces team psychological safety. These effects are not new. Earlier pre-pandemic work by Hinds and Mortensen [13] and O'Leary and Mortensen [18] describes the long-standing issues in distributed teams, including subgroup division, "us vs. them" attitude, reduced team cohesion, alienation, and weakened team identity and sense of belonging. Similarly, Olson [19] explains that remote work alters the long-term relationship between employees and their organizations, with greater autonomy coming at the cost of diminished organizational commitment. The present study builds upon related research on the relationship between remote work and attrition, and offers longitudinal evidence from a large organization, showing how resignation patterns shift across different demographic groups.

## 3. Methodology

This study employs an exploratory case study approach to investigate patterns of employee voluntary resignations before, during and after the COVID-19 pandemic. We focus on a single case study at Ericsson, a large international company headquartered in Sweden. Our goal was to examine resignation trends in HR records over time, segmented by tenure and other demographic variables, and identify emerging patterns and potential areas of vulnerability in employee retention, particularly in relation to the different work modalities: fully on-site working, fully remote working, and hybrid working.

### 3.1. The case

Ericsson is a large international telecommunication company employing over 14,500 people in Sweden, which is the center of our investigation. Ericsson develops a broad range of software- intensive products, from open-market software to complex, customized systems.

Before the pandemic, Ericsson operated as a fully on-site company. In March 2020, as most companies that can operate outside of the offices, Ericsson sent their employees to work from home. The offices reopened in September 2021. In the summer 2022, the company introduced its first work policy, requiring employees to be on-site at least 50% of the time annually. Management emphasized that in-person collaboration, peer support, and office interactions were essential for innovation, teamwork, and maintaining company culture, especially given Ericsson's reliance on agile methods. Although remote work was formally limited, enforcement was lenient: managers monitored office presence only at the building level, using average attendance as a measure of compliance. As a result, the actual office presence remained below the expectations, and Ericsson's management took the decision to raise its on-site presence mandate to 60% per week in the fall of 2024, motivated by the importance of creating and maintaining a physical hub for creative thinking, mentoring, and building strong team dynamics.

Our study is one of many studies that collects evidence of the true effects of different work modalities and provides actionable insights for organizational policy and human resource strategy.

### 3.2. Data collection and validation

This study is based on quantitative data of employment from Human Resource (HR) records and from employment exit surveys, which Ericsson sends its resignees.

### 3.2.1. Employment data

To study employee voluntary resignations, we combined two types of monthly reports:

- **Headcount reports** that list all individuals employed in a given month with their employee ID, employment form, start date, job title, city, and gender (among others), and

- **Leave reports** that contain the information about those who left the organization: employee ID, leave date, and reason (incl. resignation, retirement, lay-off, and internal mobility).

The data collection window spans from March 2016 to August 2025 with a total of 231 Excel files. Notably, the data collected originates from two different HR systems, one consequently replacing the other with two months of overlap. Further, we encountered no Leave reports for June 2019.

Exploration of the dataset surfaced several HR-specific complexities that influence the pre-processing steps and the interpretation of the raw data:

- Employment forms change over time. Besides periods of "regular employment" (which is what we study), individuals may also have held the status of being "students", "temporary workers", "subcontractors", and a dozen of other employment titles.

- Individuals can have more than one employment period (e.g., resigned or laid off, then re-hired).

- A Leave entry for a certain month does not necessarily indicate the end of an employment. HR decisions may be reverted (e.g., not laid-off after all), amended (e.g., moved to a different date), or even become obsolete (e.g., death before planned retirement). Some entries also indicate internal mobility, which are not part of our analysis.

- Attributes like an employee's city or gender may change over time.

We also encountered other issues that directly affect data quality and consistency or HR records, caused by the co-existence of two HR systems with differing data structures, subtle formatting changes over time, clerical errors in data entries, to name a few. As a result, the dataset exhibited several recurring problems:

- Employees dropping out of the Headcount without a corresponding Leave entry, sometimes just intermittently for one, two, or up to twelve months,

- Employees recorded with multiple start dates (even within the same employment period), e.g., due to transpositions like 2020-01-03 vs. 2020-03-01 or resurfacing of start dates from earlier employment periods, and

- Employees with multiple leave dates or leave reasons.

To address these issues, we applied the following data cleaning procedure:

- **Limit to regular employment**: We include only HR data on individuals who at some point had a "regular employment". This brings the number of individuals from 54,648 down to 23,377.

- **Headcount gap patching**: We filled intermittent gaps up to 12 months where employees disappeared from the Headcount and reappeared with the same form of employment and start date (i.e., no re-hiring). Among 23,377 individuals, this closed 278 gaps.

- **Split by employment period (EP)**: After filling these obvious gaps, we filtered out all non-regular EPs and were left with 23,780 regular EPs (of course, most individuals, 98%, only have one EP).

- **Start date heuristics**: For EPs that were already part of the earliest available Headcount (March 2016), we used the most frequently mentioned start date, recognizing that neither earlier nor later entries would be inherently more accurate (this affected 15,207 EPs, 64%). For all later EPs (8,570, 36%), we instead considered the first occurrence in a headcount as the start date, thus ignoring any formatting issues of start dates. For three individuals (0.01%), no start date could be determined – we exclude these individuals from all tenure-related analyses, including the survival analysis.

- **Leave date heuristics**: We assigned a single leave date per EP using the most recent value from the Leave reports, assuming that later entries are corrections and thus more correct. If no Leave record was available for an individual, we used the final month in which they appeared in the Headcount (excluding the last month in the dataset, August 2025, as this is the "last" appearance for everyone employed at that time), and categorized the leave reason as "other". We have explicit leave dates and reasons for 10,498 ended EPs (91%), and implicit leave dates for 979 (9%). Notably, this operational definition of leave date reflects the employee's effective exit from the workforce, and not the administrative moment of resignation or termination.

In the detailed investigation, we further stratified the data by defining three onboarding modalities: on-site, remote, and hybrid onboarding. Due to the absence of individual onboarding metadata in the HR system, we inferred onboarding mode based on Ericsson's work modality timelines:

- Fully on-site: Jan 2016 – Dec 2019,

- Fully remote: Jan 2020 – Aug 2021,

- Hybrid: since Sep 2021.

We considered employees with a start date between Jan 2020 – Aug 2021 to be effectively onboarded remotely, those onboarded before 2020 as onboarded onsite, and those onboarding in the hybrid work period with unclear or potentially mixed onboarding experiences, where either the new hires or their mentors might have been at least partially working remotely. Notably, we decided to include the first three months of 2020 in the remote onboarding period, because typical onboarding takes roughly half a year, and thus employees started in early 2020 are at least partially onboarded remotely.

### 3.2.2. Exit Surveys

To better understand the reasons for people quitting, we included the responses to the exit survey that Ericsson asks its resignees to take. The quarterly reports span the time period from Q3/2018 to Q3/2024 (25 Excel files in total), and contain survey ID (counted up, reflecting the time when the survey was sent out), respondents' tenure, gender, and selection of one primary and (optional) secondary reasons for leaving from a total of nine possible reasons. Each report includes responses of the previous four quarters, which extends our window back to Q4/2017.

When inspecting the data, we noticed that some survey IDs were much smaller than the rest from the same report, indicating the survey was sent out much earlier and only taken late. Based on a density plot of submission ids, we manually placed cut-off dates to separate the quarters. This back-dated 242 responses from the first report (from the "Q3/2018" report to Q4/2017 – Q2/2018) and another 10 responses from later reports by 1 to 2 quarters.

Overall, we included 1,550 responses dated between Q4/2017 and Q3/2024 matched against 3,063 recorded resignations during the same period. This yields a response rate of over 50%, which is generally considered high for organizational surveys.

### 3.3. Data analysis

The basis for all analyses is a cleaned, long-form dataset structured with one record per employee per month, covering their entire (inferred) employment period (EP). Each record contains the following fields: Month, Employee ID, Employment number, Employment status (employed, resigned, laid-off, or retired), Start date, Onboarding type (on-site, remote, or hybrid), Gender, and Tenure. Tenure is computed as the difference between the current Month and the Start Date and resets with the start of a new EP in cases of re-hiring.

This structure enables both time-series and event-history analyses at individual and aggregated levels.

### 3.3.1. Trend analysis of resignation rates

Our first analysis focused on examining monthly resignation trends across the organization. For each month, we first calculated the *voluntary resignation rate* r as the number of employees with status "resigned" divided by the total number of employees present that month. We then *annualized* that rate using the formula $1-(1-r)^{12}$. To explore differences across employee subgroups, we stratified the resignation rates by *gender* and *tenure* (see the results in Sections 4.1 and 4.2 accordingly). To reduce fluctuations and emphasize longer-term trends, all time-series plots (Figs. 1–3) were smoothed using a *Gaussian kernel* with a standard deviation of 2 months. In these plots as well as the annual summaries presented in Table 1, we excluded three anomalous months due to data quality issues:

- March 2016 and April 2016: The Leave reports during these months primarily list employees who are not present in any of the Headcount records available, resulting in unrealistically low resignation rates (approaching 0%).

- June 2019: There was no Leave report available for this month. While employees' leave date can be inferred from the Headcount, their leave reason is classified as "other", which artificially lowers the resignation rate to 0%.

### 3.3.2. Survival and hazard analysis

To assess when employees are most likely to resign, we conducted a survival analysis [5]. In the context of our study, "survival" refers to an employee choosing to remain employed, i.e., not voluntarily resigning from the company.

Our dataset for this analysis is *right-censored*. This means that we observe the full tenure only for employees who have resigned. For those still employed at the end of the observation period, or those who left for other reasons such as lay-offs or retirement, we only know the *minimum* duration of employment without a "resignation" event. To avoid left-censoring, we restricted the analysis to employees who joined the company on or after March 2016, as we lack data on prior exits.

We used the Kaplan-Meier method [15] to construct *survival curves*. These begin at 100% at the time of hire ($t = 0$) and decrease monotonically as resignations occur, representing the probability that an employee remains in the company beyond a given tenure. We compared survival curves across different employee strata using log-rank tests to assess statistically significant differences.

To identify and compare periods of elevated resignation risk (e.g., steep declines in the survival curves), we applied a piecewise Cox proportional hazard model [6]. Based on visual inspection of the survival curves, we segmented the timeline at 0.5 years and 3 years of tenure. This allowed us to estimate and compare hazard ratios across these time intervals, using the resignation risks of employees hired pre-pandemic as the baseline.

This modeling approach helped us to identify tenure-specific vulnerabilities and assess how resignation risk has evolved across the on-site, remote, and hybrid work modalities.

## 4. Results

The analysis of resignation data shows a pronounced wave of increased voluntary resignations between the summer 2021

**Table 1**
Overview of the data collected

|  | 2016 | 2017 | 2018 | 2019 | 2020 | 2021 | 2022 | 2023 | 2024 | All | Pandemic period |
|---|---|---|---|---|---|---|---|---|---|---|---|
| No of employees | 15,056 | 13,877 | 12,840 | 12,542 | 12,847 | 13,622 | 14,154 | 14,108 | 13,503 | – | – |
| Resignation rate | 4.3% | 5.3% | 4.4% | 2.7% | 1.8% | 2.8% | 4.4% | 3.2% | 2.8% | 3.5% | 2.1% |
| in females | 4.5% | 4.9% | 5.4% | 3.1% | 1.8% | 2.6% | 5.0% | 3.6% | 2.9% | 3.8% | 1.9% |
| in males | 4.3% | 5.4% | 4.1% | 2.6% | 1.8% | 2.8% | 4.2% | 3.1% | 2.8% | 3.5% | 2.1% |
| in Location 1-XL | 3.6% | 4.1% | 4.1% | 2.6% | 1.8% | 2.4% | 3.8% | 2.6% | 2.6% | 3.1% | 1.9% |
| in Location 2-L | 5.4% | 8.3% | 6.0% | 3.0% | 2.0% | 4.4% | 6.3% | 4.8% | 3.8% | 4.9% | 2.9% |
| in Location 3-L | 7.2% | 11.7% | 4.9% | 2.2% | 1.1% | 2.4% | 3.8% | 3.7% | 2.8% | 4.4% | 2.0% |
| in Location 4-L | 3.3% | 7.7% | 5.3% | 2.1% | 2.5% | 2.2% | 4.8% | 4.5% | 3.2% | 3.9% | 2.0% |
| in Location 5-M | 5.2% | 5.1% | 7.7% | 6.0% | 2.3% | 4.6% | 6.7% | 5.9% | 2.8% | 5.2% | 3.0% |
| in other locations (S) | 5.4% | 3.9% | 1.5% | 1.1% | 0.3% | 1.7% | 4.4% | 2.0% | 1.1% | 2.9% | 1.4% |

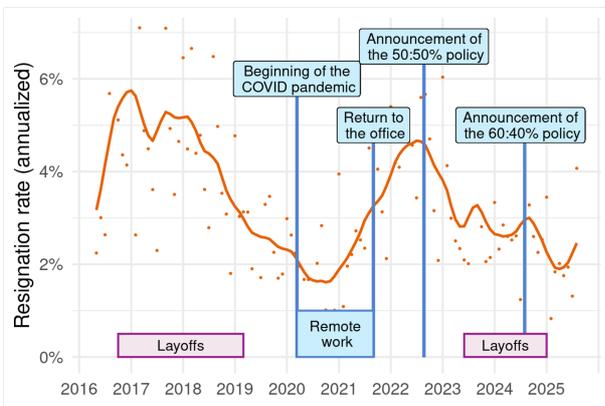

**Figure 1:** Monthly rates of voluntary resignations.

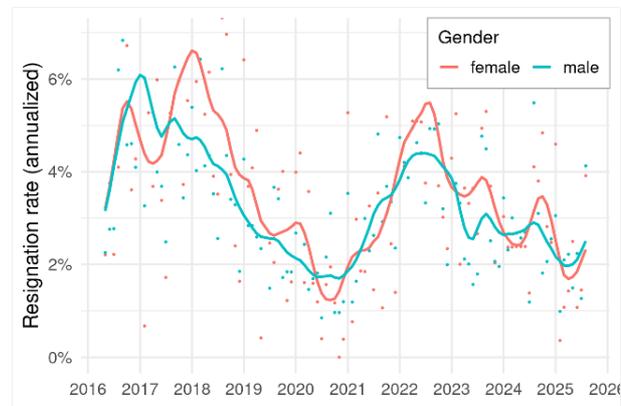

**Figure 2:** Monthly voluntary resignations by gender.

and the end of 2022 (see Fig. 1). Although not unprecedented in the history of TelCo since 2016, the post-pandemic wave stands out as the resignation rates more than doubled compared to the pre-pandemic baseline in 2019. The timing of this increase broadly coincides with the post-pandemic shift in work modalities (first return-to-office policy).

In Fig. 1, we highlight key events that may have influenced both resignation behavior and working conditions. Notably, the first visible resignation spike (2017–2018) aligns with a period of corporate layoffs. This is expected, because downsizing tends to predict voluntary resignations as "survivors" experience undermined morale and decreased organizational commitment [24, 26]. Yet, the post-pandemic wave of resignations (2021–2022) evidently *precedes* the second layoff activity, suggesting that other drivers may be at play.

To explore this further, we performed a stratified demographic analysis to better understand which employee groups were the most affected and to uncover possible reasons behind the elevated resignation rates.

### 4.1. Gender
Stratified analysis of voluntary resignations by gender (See Fig. 2) suggests that the curves for female and male employees are similar, with females resigning at slightly higher rate than males. Interestingly, this difference holds for both historical waves of resignation, in 2017–2018 and 2021–2022.

### 4.2. Tenure
Stratified analysis by the duration of employment at Ericsson indicates that employees in the tenure group of 1–2 years and 2–5 years were the most affected. This is reflected in a shift in the primary tenure group driving resignations: the likelihood of employees with 1–2 years in the company peaked after the pandemic (see Fig. 3).

One possible explanation for this shift is the decreased ability to retain people who have been onboarded remotely, i.e., those who were employed 2020 reached the tenure group of 1–2 years when the wave of resignation began, and moved to the tenure group of 2–5 years in 2022–2023.

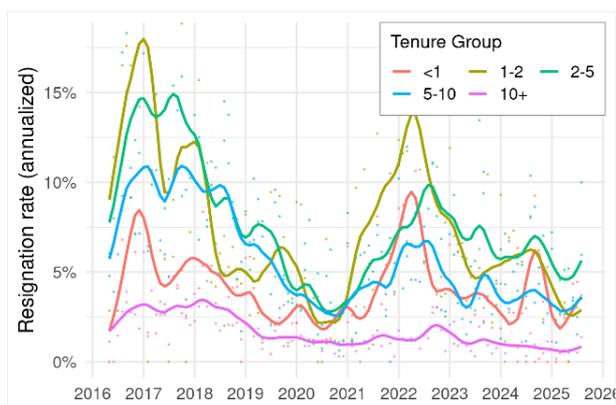

**Figure 3:** The likelihood of employees in each tenure group resigning per month.

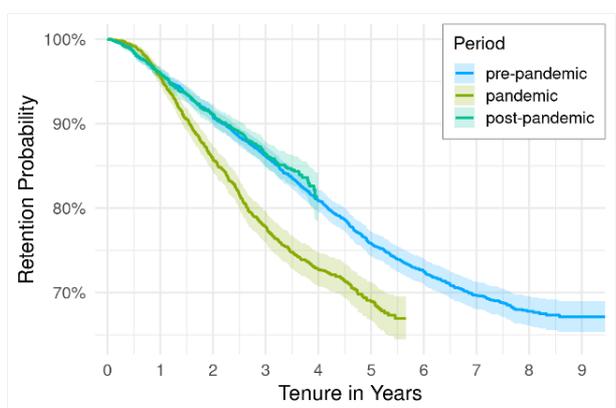

**Figure 4:** Survival curves showing the likelihood to stay in the company stratified by the onboarding modes, calculated with the Kaplan-Meier method. The shaded region represents the 95% confidence interval (uncertainty arises from individuals who did not yet leave the company and from those who left for reasons other than resignation, i.e., we could not observe their potential resignation event).

### 4.3. Remote onboarding

The differences in resignation rates across tenure segments indicated in Fig. 3 might suggest that remote onboarding fails to create a comparable sense of belonging and social integration into the company as does traditional onboarding.

We test our assumption by performing a survival analysis across the three onboarding modes (treating "resignation" as the terminating event and censoring the other reasons). A series of pairwise log-rank tests of the three survival curves in Fig. 4 shows significant differences between the pandemic and the pre-pandemic group ($\chi^2 = 38.4$, $p < 0.0001$) as well as the post-pandemic group ($\chi^2 = 47.8$, $p < 0.0001$), while the pre- and post-pandemic groups show no significant differences ($\chi^2 = 0.3$, $p = 0.6$).

Notably, the pandemic curve shows that even after Ericsson returned to in-office work, those onboarded remotely continued to resign at higher rates than before the pandemic (i.e., non-proportional hazard rates). To test this observation further, we performed a piecewise Cox proportional hazard analysis for the following three tenure periods: <0.5 years, 0.5–3 years, and 3+ years (see Fig. 5).

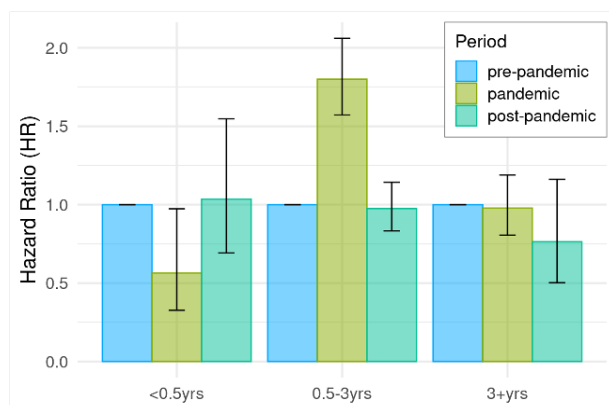

**Figure 5:** Hazard ratios of employees resigning within different intervals of their employment, stratified by onboarding period (piecewise Cox proportional hazard analysis with pre-pandemic onboarding as the baseline). Error bars indicate the 95% confidence interval.

Compared to pre-pandemic hires, those onboarded remotely have a significantly *lower* risk to resign during their first 6 months (hazard ratio HR = 0.56, $p = 0.04$). This is likely due to unwillingness to change jobs during an unprecedented cataclysm event. However, during the following 2.5 years, the risk of resignation for this group significantly *increases* (HR = 1.80, $p < 0.0001$). After three years of working in the company, the resignation likelihood stabilizes (i.e., no significant difference, HR = 0.98, $p = 0.83$).

Interestingly, pre- and post-pandemic hires show no significant differences in the first two periods (HR = 1.04, $p = 0.87$; and HR = 0.98, $p = 0.76$). Beyond 3 years of tenure, there is a hint of a tendency that post-pandemic hires are less likely to resign, but the error margins are quite wide (HR = 0.76, $p = 0.21$).

### 4.4. Reasons for resignation

To understand why employees that have been hired during the pandemic resigned at a higher rate than the new hires employed in other periods, we analyzed the reasons captured in the so-called "exit survey" that is sent to every resignee at Ericsson.

In Fig. 6, we summarize the relative count of primary reasons for resignation. At first glance, the overall top reasons to resign include Career Growth, Salary / Benefits, and Job Content. Other noteworthy contributors to attrition include Family / Personal Reasons, and Leadership / Management concerns. Interestingly, our analysis reveals that the reasons for elevated resignation activity in the two waves (2017–2018 and 2021–2022) differ.

The earlier resignation wave appears to be driven by traditional job mobility factors, such as Career Growth, Salary / Benefits and the Future and Direction of the Company.

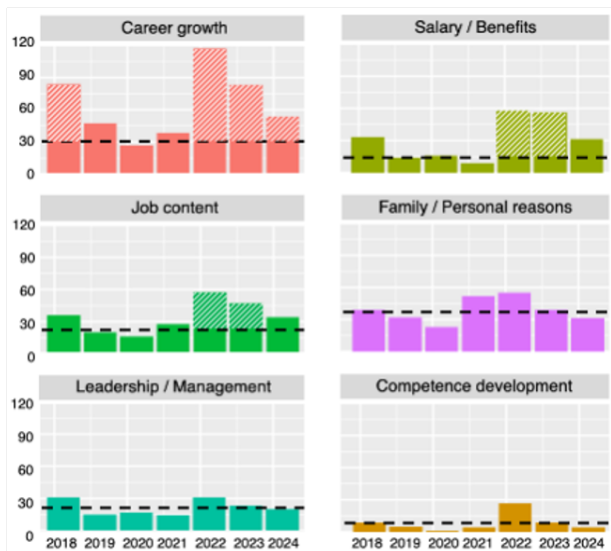

**Figure 6:** Summary of the primary reasons for resignation from the exit surveys. Highlighted areas indicate the drivers for two resignation waves of 2017–2018 and 2021–2022.

In contrast, the second wave in 2021–2022 reflects the complex interplay of pandemic-induced changes, most notably the shift to remote work and its implications for onboarding. Exit survey data reveals that Job Content reasons, typically not a huge factor, spike during this period. Combined with peaks in Career Growth and Salary / Benefits, our findings suggest that remote work may have challenged development and maintenance of meaningful roles or connections within the company, which are hallmarks of effective onboarding. All these reasons evidently drive the excess resignations.

## 5. Discussion

### 5.1. Summary of the results

Our study of voluntary resignations at Ericsson, a large company developing software-intensive systems, reveals a significant attrition wave between 2021 and 2022. The timing of this increase coincides with the company's shift from predominantly onsite work to fully remote operations in response to the COVID-19 pandemic. Stratified analysis across different demographic groups identified recent hires as the most vulnerable to resignations. These findings are consistent with national labor statistics from many countries, including the Netherlands, which report an increase in job switching during the same period, particularly among young professionals, who appear to change jobs more frequently than before.

A more detailed analysis of employees onboarded during the pandemic provides further insight into the underlying causes of this resignation wave. Employees who joined the company remotely were significantly more likely to resign, particularly within their first three years. Exit survey data reinforces this interpretation, as we find "Job Content" emerging as the most frequently cited reason for leaving among employees with short tenure during the pandemic years. This pattern suggests that remote onboarding may have led to lower engagement, reduced feedback, and limited professional development especially in the early stages of employment. These conclusions align with both pandemic and pre-pandemic studies on remote onboarding, which highlight, for example, the difficulties that remote hires face in building strong social connections with their team [21] and in navigating the steep learning curve of a new role when direct access to mentors is limited [4].

Importantly, our study provides insights into the early post-pandemic hybrid work period at Ericsson that suggest that the company has managed to reverse the negative trend and improve retention post pandemic. In the summer of 2022, Ericsson implemented a 50%-office-presence policy, followed by a 60%-office-presence policy in fall of 2024. Evidently, the employee survival curve for the post-pandemic period stabilized and returned to the pre-pandemic rate. This suggests that hybrid work that has become popular since the end of the pandemic mitigates the challenges observed during the fully remote mode.

Our findings align with and help explain the paradox of remote work described in prior research. While remote work is often associated with higher job satisfaction, it may simultaneously undermine long-term retention. For example, de Souza Santos and Ralph [23] documented how remote-first and hybrid teams suffer from reduced coordination, limited interaction, and eventual detachment among team members. Olson [19] similarly argued that increased autonomy in remote settings may come at the cost of diminished organizational commitment. Our results align with these theories and extend them by showing that remote onboarding, in particular, exacerbates these effects.

Additionally, our findings help contextualize Golden's [12] observation of an inverted U-shaped relationship between remote work and job satisfaction. Remote onboarding at Ericsson appears to fall on the declining side of this curve, where the flexibility and autonomy of remote work are overshadowed by isolation, weak peer connections, and lack of integration into the organizational culture. These conditions may have driven newly hired employees to seek alternative employment once pandemic-related uncertainty eased.

In summary, our study supports and empirically extends prior research by showing that remote work may introduce structural and social vulnerabilities, particularly during onboarding, and how these result in increased attrition. Further research is needed to evaluate the long-term success of hybrid onboarding and to understand how team-level practices shape these outcomes. In particular, attention should be paid to team work modalities, as prior research has shown that even a single remote team member can reduce feedback exchange and hinder learning across the team [8].

A Wave of Resignations in the Aftermath of Remote Onboarding## 5.2. Alternative explanations

It is fair to question whether the failure of remote onboarding alone accounts for the increase in resignations among recent hires at Ericsson during 2021–2022. Several alternative explanations merit consideration.

**Generational differences**. One hypothesis is that newer entrants to the labor market, particularly younger generations, demonstrate different workplace and job expectations and a lower threshold for switching employers. However, this explanation appears insufficient. If generational behavior were the primary driver, we would expect the elevated resignation trend to persist or even increase over time. Instead, we observe a sharp decline in resignations after 2022.

**Labor market dynamics**. A second explanation is related to broader labor market conditions. For example, a surge in available job opportunities, especially in tech and IT, could have made it easier and more attractive for employees to switch jobs during that period. National statistics for the Information and Communication Technology sector in Sweden[5] shows a marked drop in job vacancies in Q2/2020, consistent with the uncertainty during the early pandemic. This was followed by a strong rebound beginning in Q2/2021 with a peak in vacancies between Q4/2021 and Q3/2022, likely reflecting a catch-up in delayed hiring. Such a surge in demand may have created a "resignation window", particularly for employees dissatisfied with their roles or onboarding experiences. However, the subsequent period of improved retention complicates the labor market hypothesis. Although job vacancies slightly declined between Q3/2023 and Q1/2024, the period of stabilization in resignations at Ericsson begins before the decline in job openings. This temporal mismatch suggests that job availability alone cannot fully explain the observed patterns. In fact, it may be that resignation behavior helped trigger the job availability spike, as organizations attempted to backfill roles during the 2021–2022 attrition wave.

**Macroeconomic uncertainty and risk aversion**. A third factor concerns the broader economic climate in the late 2022 and 2023 characterized by high inflation, sharply rising interest rates, and subsequent cooling in the labor market as hiring froze. Such conditions of economic uncertainty are known to foster risk aversion and financial caution. The Swedish inflation peaked at 12.3% in Q4/2022 and steadily declined to 1.6% in Q3/2024. The recession concerns period is a plausible explanation for the risk-averse behavior among employees, discouraging job mobility, as the timing corresponds with the observed stabilization in resignation rates at Ericsson. However, if macroeconomic uncertainty were the primary driver, we would expect renewed resignation activity once inflation and interest rates normalized in 2024. Yet the reverse trend is not evident in the data, suggesting that while economic uncertainty may have contributed to improved retention, it does not fully explain the preceding increase in resignations – nor the lack of resurgence afterward.

## 6. Conclusions

In this exploratory case study, we demonstrate how remote onboarding during the pandemic is associated with higher resignation rates in the early years of employment. Our findings suggest that onboarding modality significantly influences employees' long-term attachment to the organization, with remote onboarding potentially weakening social integration and organizational commitment.

Our results have practical implications for HR leaders and policymakers seeking to design effective hybrid work environments. Organizations should critically assess how they onboard new hires in remote or hybrid contexts, emphasizing opportunities for relationship building, mentoring, and cultural integration. While return-to-office (RTO) mandates are looked at as a threat to job satisfaction, it is equally important to understand that fully remote work may be a threat to job satisfaction too, as already demonstrated in related studies from before the pandemic [12]. We suggest that differentiated RTO mandates can be applied as a remedy to establish workplace cohesion, at the same time emphasizing that these shall be equally applied to new hires and their peers and mentors in the team. We thus recommend companies to promote increased onsite presence for new hires and co-presence for teams that onboard new members.

Future research should further investigate the causal mechanisms, and quantify the effect of the RTO mandates, as early studies suggest increasing resignations following RTO mandates with more senior employees leaving at higher rates [27].

## Acknowledgements

We are grateful to our contacts at Ericsson for their active engagement in our research. Their continuous collaboration and willingness to put effort into making corporate data accessible were essential to making this study possible.

## Funding

This study was funded by the Swedish Knowledge Foundation through the projects S.E.R.T. (grant number 2018/010) and WorkFlex (grant number 2022/0047).

## Author contribution declaration

**Darja Smite**: Conception, Methodology, Formal analysis, Writing – original draft, Data visualization. **Franz Zieris**: Conception, Methodology, Data curation, Formal analysis, Writing – review and editing, Data visualization. **Lars-Ola Damm**: Conception, Writing – review and editing.## Data availability statement

The policies collected during the study and the HR data and exit surveys are not available for confidentiality reasons.

---

[5] Statistikmyndigheten SCB, Job vacancies by quarter and industry (2015-2024): https://www.statistikdatabasen.scb.se/pxweb/sv/ssd/START_AM__AM0701__AM0701A/KV15LedigajobbVak07/table/tableViewLayout1/

Smite et al.: *Preprint submitted to Elsevier*     Page 8 of 9

## Conflict of interest statement

The authors have no competing interests to declare.